\documentclass[prl,aps,twocolumn,superscriptaddress]{revtex4-1}
\usepackage{graphicx,color}
\usepackage{amsthm}
\usepackage{amsfonts}
\usepackage{algorithmic}
\usepackage{enumerate}
\usepackage{latexsym}
\usepackage{amsmath}
\usepackage{amssymb}
\usepackage[colorlinks=true,citecolor=blue,linkcolor=blue]{hyperref}

\emergencystretch=\maxdimen
\def\degree{${}^{\circ}$}

\begin{document}

\title{Superconductivity in twisted multilayer graphene: a smoking gun in recent condensed matter physics}

\author{Chu Yonghuan}
\affiliation{Department of Physics, Beijing Normal University, Beijing 100875, China\\}

\author{Zhu Fangduo } 
\affiliation{Department of Physics, Beijing Normal University, Beijing 100875, China\\}

\author{Wen Lingzhi} 
\affiliation{Department of Physics, Beijing Normal University, Beijing 100875, China\\}

\author{Chen Wanying} 
\affiliation{Department of Physics, Beijing Normal University, Beijing 100875, China\\}

\author{Qiaoni Chen} 
\affiliation{Department of Physics, Beijing Normal University, Beijing 100875, China\\}

\author{Ma Tianxing} 
\email{txma@bnu.edu.cn}
\affiliation{Department of Physics, Beijing Normal University, Beijing 100875, China\\}

\begin{abstract}
In this article, we review the recent discoveries of exotic phenomena in graphene, especially superconductivity. It has been theoretically suggested for more than one decade that superconductivity may emerge in doped graphene-based materials. For single-layer pristine graphene, there are theoretical predictions that spin-singlet $d+id$ pairing superconductivity is present when the filling is around the Dirac point. If the Fermi level is doped to the Van Hove singularity where the density of states diverges, then unconventional superconductivity with other pairing symmetry would appear. However, the experimental perspective was a bit disappointing. Despite extensive experimental efforts, superconductivity was not found in monolayer graphene. Recently, unconventional superconductivity was found in ``magic-angle" twisted bilayer graphene. Superconductivity was also found in ABC stacked trilayer graphene and other systems. In this article, we review the unique properties of superconducting states in graphene, experimentally controlling the superconductivity in twisted bilayer graphene, as well as a gate-tunable Mott insulator, and the superconductivity in trilayer graphene. These discoveries have attracted the attention of a large number of physicists. The study of the electronic correlated states in twisted multilayer graphene serves as a smoking gun in recent condensed matter physics.


\end{abstract}

\maketitle

\section{Introduction}
Graphene is a single-layer material in which the carbon atoms are arranged in a hexagonal structure.
It is one of the most exciting systems and has attracted a large amount of research interests in the past decades\cite{PhysRev.71.622,RevModPhys.81.109}.
One of the novel properties of graphene is that its low-energy excitations are massless, chiral, Dirac fermions\cite{Novoselov2005Two,Zhang2005Experimental}. Another interesting feature of graphene is the possible strong interaction and correlation effects,
which cause many exotic states. Among them, superconductivity has been intensively studied.

The chemical potential of graphene can be tuned through an electric field effect\cite{Novoselov2005Two}; hence, it is possible to introduce charge carriers, electrons, or holes into graphene. This paves the way for realizing superconductivity, and various theoretical works have been conducted to discuss the onset of superconductivity and the pairing symmetry. Due to the special structure of the honeycomb lattice, a mean field study suggests that an extended-s (ES) SC phase is present in doped graphene\cite{Uchoa2007Superconducting}.
Further, many more studies indicate that the pairing symmetry is $d_{x^2-y^2}+id_{xy}$ ($d+id$) when the Fermi energy is quite close to the Dirac point. These studies include a weak-coupling functional renormalization group study\cite{Honerkamp2008Density}, a mean field study of a phenomenological Hamiltonian\cite{Baskaran2002Resonating}, variational Monte Carlo simulations\cite{Pathak2010Possible}, and the determinant quantum Monte Carlo method\cite{PhysRevB.84.121410,Lin2015Quantum}. It has been suggested that the spin-singlet $d+id$-wave state is the chiral $d$-wave superconducting state, which is characterized by the breaking of both time-reversal and parity symmetries\cite{Black_Schaffer_2014}. However, a constrained-path quantum Monte Carlo study does not support the presence of intrinsic superconductivity in single-layer graphene. The study is based on scaling analysis of the ground state pairing correlation. The long-range part of the $d+id$ pairing correlation decreases as the lattice size increases and tends to vanish in the thermodynamic limit\cite{PhysRevB.84.121410}. This indicates that the electron correlation in lightly doped graphene is not strong enough to produce an intrinsic superconductivity because of the low density of states (DOS). In graphene, the DOS near the Fermi level is almost zero when the filling is close to the Dirac point. Therefore, relatively weak or intermediate short-range repulsion interactions may not induce any phase transitions at low temperatures\cite{RevModPhys.81.109}. It has been recently found that near the type-I Van Hove singularity (VHS) when the filling is $1/4$, there are singular phases, such as spin density waves\cite{PhysRevLett.101.156402,Li2012Spontaneous}, $d+id$\cite{PhysRevB.85.035414,PhysRevB.86.020507,Nandkishore2012Chiral,PhysRevB.78.205431,PhysRevB.75.134512,PhysRevB.81.085431}, topological superconductivity\cite{RevModPhys.82.3045,RevModPhys.83.1057} and Chern-band insulators\cite{PhysRevB.99.155415}. In these phases, the DOS at the Fermi level diverges logarithmically. Such a logarithmically diverging DOS close to a VHS may significantly raise the superconducting transition temperature\cite{Ma2014}.
A renormalization group analysis shows that the topological triplet $p+ip$ superconductivity can generally occur in a type-II VHS system\cite{PhysRevB.92.035132,PhysRevB.92.174503},
similar with the results of quantum Monte Carlo studies\cite{Ma2014,Ma2015}. In the type-II VHS system, the saddle points are not at points of time-reversal-invariant momenta. Although there are many theoretical predictions, reliable experimental studies remain rare. Actually, the main difficulty lies in the realization of high doping levels through chemical approaches. Thus, the occurrence of intrinsic superconductivity in graphene is still a controversial subject.

\begin{figure}[tbp]
\includegraphics[scale=0.5]{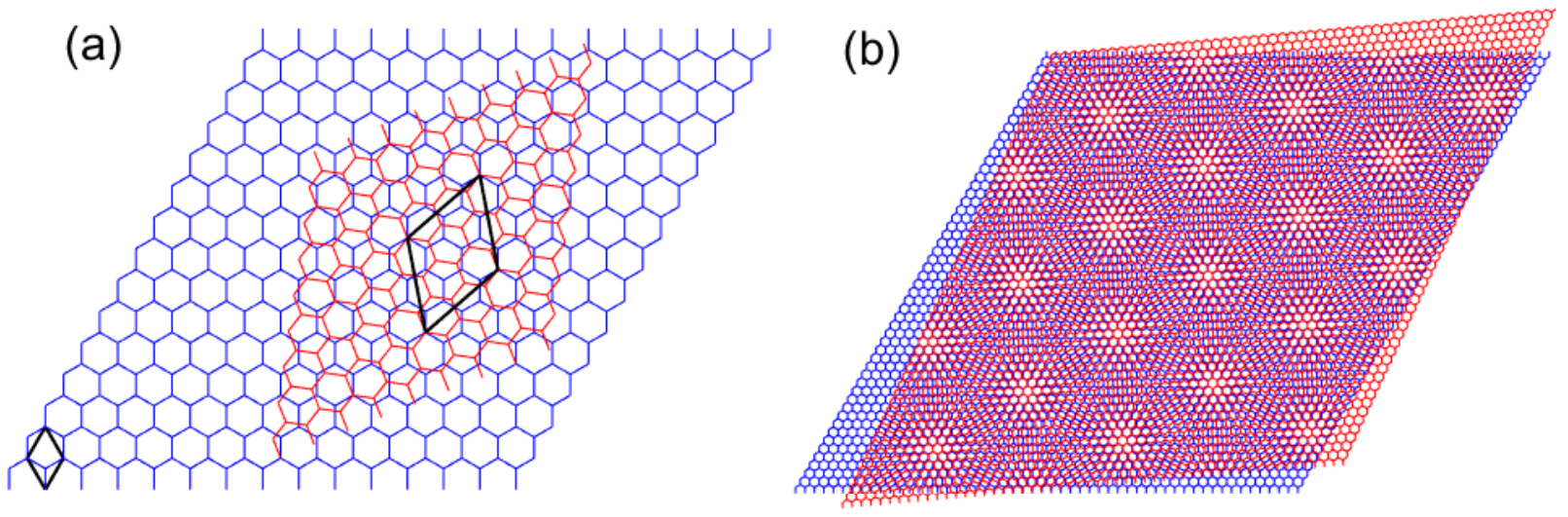}
\caption{(a) When a graphene bilayer is twisted so that the top sheet is rotated out of alignment with the lower sheet, the unit cell (the smallest repeating unit of the material's 2D lattice) becomes enlarged; (b) for small rotation angles, a moir\'{e} pattern is produced in which the local stacking arrangement varies periodically.
} \label{structure-}
\end{figure}

Exciting progress has been recently made: on March $5$ of $2018$, two papers in $Nature$ reported the discovery of novel electronic ground states in twisted bilayer graphene (TBG)\cite{Cao2018A,Cao2018B}. The structure of TBG is displayed in Fig.\ref{structure-}. Upon rotating the layers away from Bernal stacking to the so-called ``magic angle" of approximately 1.1\degree, the interplay between the resulting moir\'{e} superlattice and hybridization between the layers leads to the formation of an isolated flat band at the charge neutrality point (CNP). Near this flat band angle, one of the novel states has been interpreted as the Mott insulator at half band filling. As charge carriers are introduced to TBG, which drives the ``Mott insulator" away from half band filling, a superconducting phase appears. This behaviour is quite similar to that occurring in cuprates and other strongly correlated materials. Since the DOS is also rather local, it is believed that there are strongly correlated physics in TBG. This discovery has inspired a huge number of physicists to study TBG.

The magic angle in TBG is governed by the interlayer coupling strength and the individual layer band structure.
To elucidate the correlated electronic states, the magic angle is controlled by continuous tuning of the superlattice band structure\cite{Liu2019Spin}. 
Quantum geometry, namely, a finite quantum metric contributes to the superconductivity in TBG, due to the flat band nature of the system\cite{Julku101}. 
Because of the electron-electron interactions, magic angle TBG displays unusual spectral characteristics, which have been observed by high-resolution STM\cite{Xie2019Spectroscopic}. The discovery of tunable unconventional superconductivity and Mott physics in TBG has inspired research on the ABC trilayer graphene/hexagonal boron nitride (hBN) superlattice and twisted double bilayer graphene (TDBG). An electric field can be used to adjust the correlated insulator gaps\cite{10.1038/s41567-018-0387-2}, and this provides the possibility of adjusting the moir\'{e} flat band.

The outline of this paper is as follows. In Sec. II, we concentrate on the correlated insulating phases and the possibility of superconductivity in TBG, from both experimental and theoretical perspectives. In Sec. III, we focus on the electronic properties of trilayer graphene and discuss the recent evidence of the Mott insulating state and superconducting state in the ABC-stacked graphene/hBN heterostructure. Finally, in Sec. IV, we summarize the discussions and present our vision for future research.

\section{Twisted bilayer graphene}

\begin{figure}[tbp]
\includegraphics[scale=0.15]{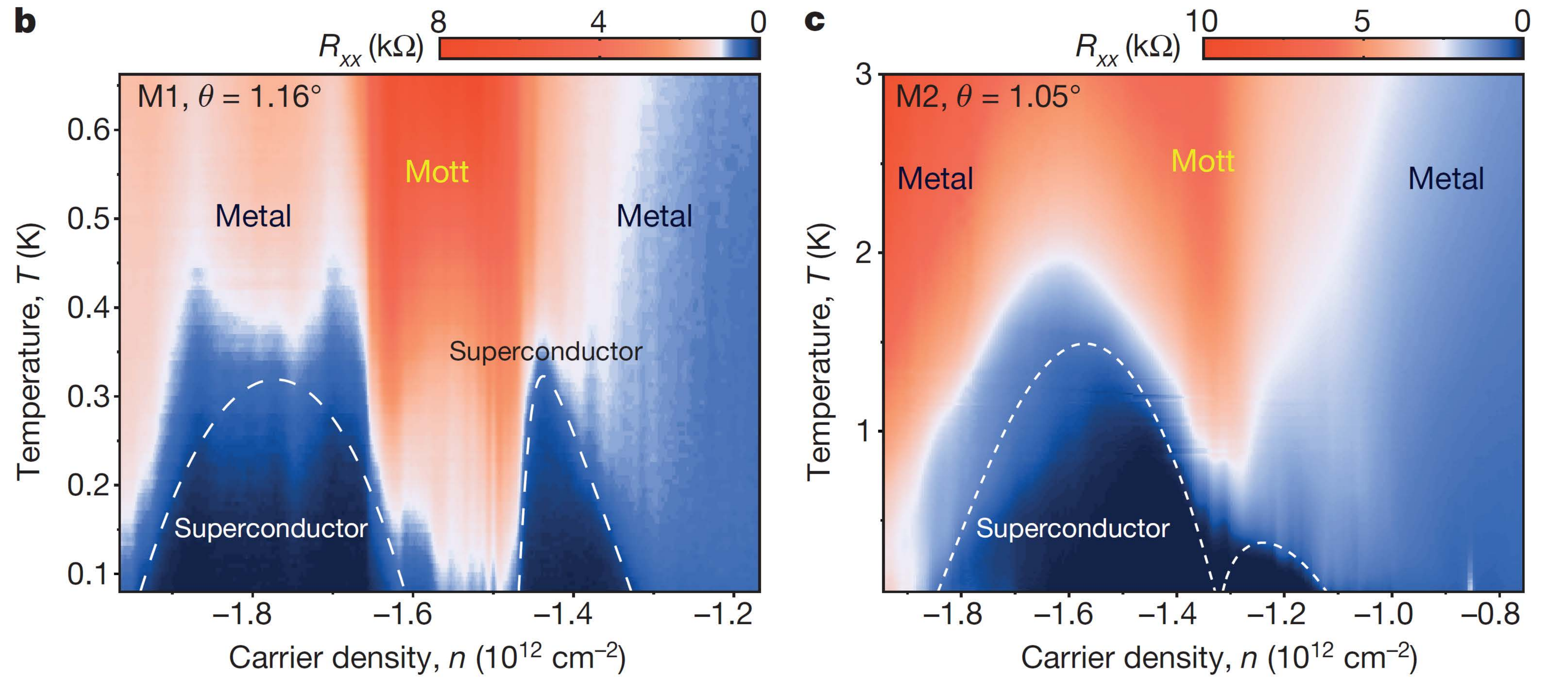}
\caption{(b) Two superconducting domes are observed next to the half-filling state, which is labelled Mott and centred around $-n_s/2 = -1.58 \times 10^{12} cm^{-2}$. The remaining regions in the diagram are labelled as Metal owing to the metallic temperature dependence. The highest critical temperature observed is $T_c = 0.5 K$ (at 50$\%$ of the normal-state resistance). (c) As in b, two asymmetric and overlapping domes are shown. The highest critical temperature is $T_c = 1.7 K$. Reproduced with permission from Ref.\cite{Cao2018B}.
} \label{Yuan Cao}
\end{figure}

The fabulous experiment by Y. Cao et al.\cite{Cao2018A} showed that the DOS near the Fermi level of magic angle TBG is approximately three orders of magnitude greater than that of two uncoupled graphene sheets. The reasons for this are that the Fermi velocity is decreased and the electron localization increases. This suggests that the electrons in magic angle TBG become strongly correlated. More exciting results were found by the same group\cite{Cao2018B}. Superconductivity occurs when the Fermi energy $E_F$ is tuned away from half band filling. The doping-dependent phase diagram is quite similar to that of doped cuprates and is displayed in Fig.\ref{Yuan Cao}. It is also asymmetric about half band filling with hole and electron doping.
Similar to other superconductors, a magnetic field suppresses the superconductivity in TBG, consistent with the Meissner effect and Ginzburg-Landau theory.
The conductance-doping relation reveals a correlated insulating state at the half-filled flat band.
It was further shown that the insulating states at half filling should be destroyed by both temperature and magnetic field.
Y. Cao and collaborators claimed that the correlated insulating state when the lower flat band is at half band filling is in a Mott-insulating phase.
Whether it is a Mott insulator or not is still disputed.
The experimental results indicate that the ``Mott gap" is approximately 0.31 meV, much less than the band width of W$\approx$ 10 meV.
The Mott gap can be ablated by a magnetic field larger than $B$=6T.
Due to the asymmetric doping dependence of the effective mass, the resistances are not Mott-like.
The energy scale between superconductivity and Mott is also quite close.
Based on the above facts, some researchers believed that it is inappropriate to regard the ``correlated insulating" phase as a parent of the superconducting phase, and the results from the weak-coupling RPA approach indicate that the correlated insulating state at half filling might be some kind of density wave (DW) instead of a Mott insulator\cite{PhysRevLett.121.217001}.

By synthesizing more pure samples, M. Yankowitz \textit{et al.}\cite{Yankowitz1059} also found superconductivity in TBG. In their experiments, by fixing the twist angle to 1.1\degree, an isolated low-energy flat band was introduced into double-layer graphene\cite{PhysRevX.9.031021}. This led to the appearance of gate-tunable superconducting and correlated insulating phases. In addition to the twist angle, the interlayer coupling could also be tuned to discover novel superconducting and insulating phases. By controlling the hydrostatic pressure to control the interlayer coupling, M. Yankowitz \textit{et al.} found superconductivity at a twist angle larger than 1.1\degree. Hydrostatic pressure also induced Metal-Insulator transition in TBG\cite{Padhi99}. And adding an insulating tungsten diselenide (WSe$_{2}$) monolayer between the hexagonal boron nitride(hBN) and the TBG stabilizes superconductivity at twist angles much smaller than the magic angle\cite{Arora583}. These experimental results again demonstrate that TBG is a novel type of tunable platform for exploring correlated states.

These experimental results raise several important questions: Is the correlated insulator a Mott insulator or not? What is the pairing mechanism of the superconductivity? What is the superconducting pairing symmetry? To answer these questions, numerous different models and methods have been employed for TBG. We present a brief summary of these works below.

Many theories were proposed to explain the unconventional superconductivity and the correlated insulating states\cite{PhysRevX.8.031089,PhysRevX.9.031021,PhysRevX.8.031089,PhysRevX.8.041041,PhysRevX.8.041041,PhysRevLett.121.217001,PhysRevLett.121.087001,PhysRevLett.121.257001,PhysRevLett.122.257002,PhysRevB.99.144507,PhysRevB.99.195114,PhysRevB.99.121407,PhysRevB.99.094521,82,26,
Esquinazi100,Volovik107,PhysRevB.100.115135,Huang2019Antiferromagnetically,PhysRevB.91.134514,Chen2019Tunable,PhysRevB.98.214521,Kennes98,PhysRevLett.122.026801,You2019Superconductivity,Sharma2,Lu574,Talantsev10,Scheurer2,Samajdar102,PhysRevB.98.045103,PhysRevX.8.031088,
PhysRevX.8.031087,PhysRevB.99.195455,PhysRevResearch.1.033072,PhysRevLett.122.246401,PhysRevResearch.1.033126,PhysRevB.100.045111,PhysRevB.98.121406,PhysRevLett.124.097601,Liu2019Correlated,Liu2019Nematic,Bultinck2019Ground,PhysRevX.8.031088,PhysRevX.8.031087}. The starting point of a theoretical work is to construct an effective model, and there are mainly two different routes to construct effective models for TBG. One is to construct the tight-binding model from local Wannier orbitals\cite{PhysRevB.98.045103,PhysRevX.8.031088,PhysRevX.8.031087,PhysRevB.99.195455,PhysRevResearch.1.033072}, and the other is to construct the continuum model from the scattering between Dirac points that belong to different layers\cite{PhysRevB.86.155449}.

TBG is obtained by rotating two layers of graphene at an angle $\theta$. In the direction parallel to the layer, the atoms form a superlattice structure at specific values of the twist angle $\theta$\cite{PhysRevB.90.155451}. When the twist angle $\theta$ is small, the superlattice constant of the moir\'{e} pattern is $L=\sqrt{3}d/(2\sin\frac{\theta}{2})$, where $d$ is the length of the $C-C$ bond\cite{PhysRevB.78.113407,PhysRevB.77.235403,PhysRevLett.100.125504,PhysRevB.77.165415,PhysRevLett.106.126802,PhysRevLett.109.186807,PhysRevB.83.205403}.
When the twist angle is $\theta_e \sim 5$\degree, the Fermi speed $v_F$ rapidly decreases to zero, and the Fermi energy level $E_F$ is within the flat bands\cite{PhysRevB.90.155451}.

In the tight-binding representation, the Hamiltonian for TBG is written as\cite{PhysRevB.96.155416,PhysRevLett.99.256802,PhysRevB.82.121407}
\begin{eqnarray}
H=&-&t\sum_{\mathbf{l}\left\langle i,j\right\rangle \sigma}(a^{\dagger}_{\mathbf{l}i\sigma}b_{\mathbf{l}j\sigma}+b^{\dagger}_{\mathbf{l}i\sigma}a_{\mathbf{l}j\sigma})\nonumber\\
&-&\sum_{i,j,\mathbf{l}\neq \mathbf{l}'\sigma}t_{ij}(a^{\dagger}_{\mathbf{l}i\sigma}a_{\mathbf{l}'j\sigma}+a^{\dagger}_{\mathbf{l}i\sigma}b_{\mathbf{l}'j\sigma}+b^{\dagger}_{\mathbf{l}i\sigma}a_{\mathbf{l}'j\sigma}+b^{\dagger}_{\mathbf{l}i\sigma}b_{\mathbf{l}'i\sigma}) \nonumber\\
&+&\mu\sum_{i,\mathbf{l},\sigma}(a^{\dagger}_{\mathbf{l}i\sigma}a_{\mathbf{l}i\sigma}+b^{\dagger}_{\mathbf{l}i\sigma}b_{\mathbf{l}i\sigma}) \nonumber\\
&+&U\sum_{i,\mathbf{l}}(n_{\mathbf{l}ai\uparrow}n_{\mathbf{l}ai\downarrow}+n_{\mathbf{l}bi\uparrow}n_{\mathbf{l}bi\downarrow})
+V\sum_{\mathbf{l}\langle i,j\rangle }(n_{a\mathbf{l}i}n_{b\mathbf{l}j}),
\label{model}
\end{eqnarray}
Here, $a_{\mathbf{l}i\sigma}$ ($a_{\mathbf{l}i\sigma}^{\dag}$) are annihilation (creation) operators acting at site $\mathbf{r}^{a}_{\mathbf{l}i}$ of layer $\mathbf{l}$ ($\mathbf{l}=1,2$) with spin $\sigma$ ($\sigma$=$\uparrow,\downarrow$) on sublattice A, and $n_{\mathbf{l}ai\sigma}=a_{\mathbf{l}i\sigma}^{\dagger}a_{\mathbf{l}i\sigma}$ is the electron number operator. Likewise, $b_{\mathbf{l}i\sigma}$ ($b_{\mathbf{l}i\sigma}^{\dag}$) and $n_{\mathbf{l}bi\sigma}=b_{\mathbf{l}i\sigma}^{\dagger}b_{\mathbf{l}i\sigma}$ are annihilation (creation) and electron number operators at site $\mathbf{r}^{b}_{\mathbf{l}i}$ on sublattice B. $t$ is the hopping integral between the nearest neighbours within the same layer, with $t\approx 2.7eV$ from the estimation of ab initio calculations. $\mu$ is the chemical potential, while $U$ and $V$ denote the onsite Coulomb repulsion and the nearest neighbour Coulomb interaction. $t_{ij}$ is the interlayer hopping integral between sites $\mathbf{r}_{1i}$ and $\mathbf{r}_{2j}$, and the definition is
\begin{eqnarray}
t_{ij}=t_p e^{-[(|\mathbf{r}^{d}_{1i}-\mathbf{r}^{d'}_{2j}|)-d_0]/\xi},
\label{tc}
\end{eqnarray}
where $t_p=-0.17t$, the perpendicular interlayer distance $d_0=0.335$ nm, and $\xi=0.0453$ nm\cite{PhysRevB.96.155416}.

Based on this tight-binding model, Huang \textit{et al.}\cite{Huang2019Antiferromagnetically} studied the phase transition between different phases by using determinant Monte Carlo method. They found that there is a Mott insulator transition at $U_c\simeq 3.6$ at the CNP, and $d+id$ superconductivity pairing dominates at finite doping.
The shortcoming of their simulation was that the lattice size they could simulate was rather small. The unconventional superconductivity reported in TBG stimulated further research on electronic correlated phenomena.

To capture the low-energy physics of magic angle TBG\cite{PhysRevX.8.031089,PhysRevB.98.045103}, Fan Yang's group\cite{PhysRevLett.121.217001} proposed a honeycomb lattice model based on the characteristics of the flat bands at three high symmetry points $\Gamma$, $K$, and $M$. The two sublattices of the honeycomb lattice represent the atoms in the two layers, and the Wannier orbitals at each site are the $p_x$ and $p_y$ orbitals. The hopping integrals of the $p_{x,y}$ orbitals, which represent the coexisting $\sigma$ and $\pi$ bondings\cite{PhysRevB.77.235107,PhysRevLett.101.186807,PhysRevB.90.075114,PhysRevB.90.085431,PhysRevB.91.134514}, are constructed via the Slater-Koster formalism by symmetry analysis. Via random-phase-approximation-based calculations, they found chiral $d+id$ topological superconductivity bordering the correlated insulating state near half filling, identified as the non-coplanar chiral spin-density wave (SDW) ordered state, featuring the quantum anomalous Hall effect\cite{Liu2019Anomalous}. Liang Fu's group proposed a two-orbital Hubbard model on an emergent honeycomb lattice\cite{PhysRevB.98.045103,PhysRevB.98.079901}. The model was constructed from Wannier orbitals that extend over the size of supercells by considering the electronic structure of narrow minibands and the effect of the Coulomb interaction.

From the viewpoint of weak coupling, there are some similarities between the models of Yang's group and Fu's group. In both works, they found that the superconducting pairing mechanism is DW fluctuations and that the instability is driven by the Van Hove singularity and Fermi-surface nesting. Several other groups also support this point\cite{You2019Superconductivity,PhysRevLett.122.026801,PhysRevB.98.205151,PhysRevB.98.214521}. The differences in these works are the degeneracy of SDWs or charge density waves (CDWs), singlet pairing or triplet pairing, and CDWs vs. SDWs\cite{PhysRevB.98.045103,You2019Superconductivity,PhysRevLett.121.217001}. There is also another possibility: the SDW and CDW orders can be mixed, and a novel chiral SO(4) spin-charge DW state with exotic properties has been argued\cite{82}. The model of Fu's group has also been verified by explicit numerical calculations\cite{PhysRevX.8.031087,PhysRevX.8.031088}. With the finite temperature determinant quantum Monte Carlo method, our group studied Fu's model\cite{Chen2019Tunable}. From the temperature dependence of the conductance, we found the Mott insulator transition, and the Mott phase on the strong coupling side is accompanied by antiferromagnetic long-range order. Over a wide filling region, a $d+id$ superconducting phase is present because of the correlations between electrons. Dante \textit{et al.} also find a crossover between $d+id$ superconductivity and antiferromagnetic insulating behavior near half filling of the lowest electron band when the temperature is increased,
in which an unbiased renormalization group approach are used to predict SC in TBG\cite{Kennes98}. 

There are many other more results on the superconductivity in TBG by different methods and models. Sharma \textit{et al.} used the Migdal-Eliashberg framework on a one-parameter effective lattice model for TBG  show that a superconducting state can be achieved by means of collective electronic modes in TBG\cite{Sharma2}. Talantsev \textit{et al.} used existing BCS theory and Ginzburg-Landau(GL) models to analyze experimental data from magic angle TBG reported by Cao \textit{et al.} and Lu \textit{et al.}\cite{Lu574}, and they found that magic angle TBG is a moderately strong coupled two-band superconductor with $s-$ or $p-$wave symmetry\cite{Talantsev10}. By applying the strong-coupling Eliashberg theory with both inter- and intraband quantum critical pairing interactions, a nodeless $s^{\pm}$ wave pairing symmetry has been discussed.
 Even though it is distinctly different from previous theoretical proposals, it also highlights the multigap nature of the superconductivity and places TBG in the same classes as iron pnictide, electron-doped cuprate, and some heavy fermion superconductors\cite{26}.

Besides twisted bilayer graphene, the family of moir\'{e} superlattice systems displaying correlated
physics expands rapidly\cite{1Physics12MacDonald}. Another one such
heterostructure that has generated much interest is twisted
double-bilayer graphene (TDBG)\cite{2NaturePhysics16Shen,3Nature583Liu,4Nature583Cao,5PhysRevLett.123.197702}, in which two AB stacked
graphene bilayers are twisted relative to each other. In the pairing of the twisted double-bilayer graphene and the related moir\'{e} superlattice systems, Scheurer \textit{et al.} separately discussed singlet and triplet pairing. The behavior of these two channels is close to the invariant limit when the system spins independently in two valleys, realizing SU(2)$_{+} \times$ SU(2)$_{-}$ symmetry\cite{Scheurer2}. They also show that regardless of the microscopic details, triplet pairing will be stabilized if the collective electronic fluctuations breaking the enhanced SU(2)$_{+} \times$ SU(2)$_{-}$ spin symmetry of these systems are odd under time reversal, even if the main SU(2)$_{+} \times$ SU(2)$_{-}$ symmetrical part of the pairing glue is provided by phonons\cite{Samajdar102}. 

\section{Trilayer graphene }
\begin{figure}[h]
	\centering
	\includegraphics[scale=0.55]{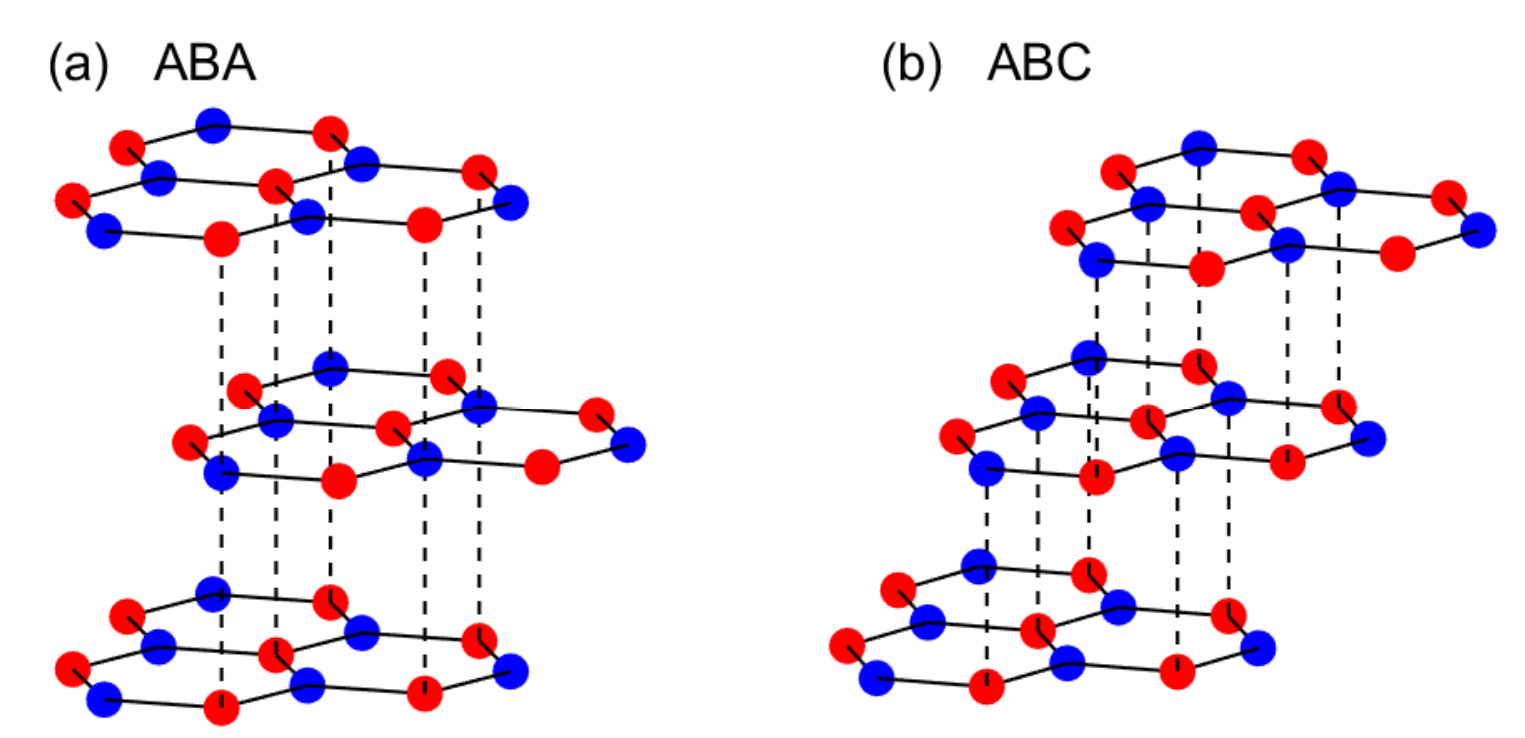}
	\caption{Crystal structure of ABA (a) and ABC (b) trilayer graphene.}
	\label{structure}
	\end{figure}

The trilayer graphene (TLG) system also provides an intriguing platform for the study of unconventional superconductivity and strongly correlated electron physics and has drawn a huge amount of attention in the last decades.
TLG has two types of naturally stable structures, and they are shown in Fig. \ref{structure}. In the ABA stacking structure, the atoms of the middle layer are staggered with the atoms of the other two layers, and the atoms of the top layer align just above the atoms of the bottom layer. In the ABC stacking structure, the atoms of different layers are staggered. In both structures, when two neighbouring layers are staggered, the atoms belonging to one of the sublattices of the higher layer are on top of the atoms belonging to one of the sublattices of the lower layer. The atoms belonging to the other sublattice of the higher layer are within the hexagons formed by the lower layer.

This distinction in the stacking order causes dramatic differences in the band structures. Several groups calculated the low-energy band structures of different stacking orders, such as the work by Aoki M and Amawashi H in 2007\cite{Aoki2007}. It is well known that the ABA-stacked TLG (ABA-TLG) has a semimetallic property since there is a small band overlap at the Dirac point. The linear part of its band structure is similar to that of monolayer graphene\cite{PhysRevB.73.144427,PhysRevB.73.245426}, and the parabolic part is similar to that of bilayer graphene\cite{PhysRevLett.97.036803,PhysRevB.75.193402,PhysRevB.77.115313,Aoki2007}. The ABC-TLG behaves like a semiconductor since there is a band gap ($\sim$ 20 meV) near the Dirac point. ABC-TLG was predicted to be a tunable narrow-gap semiconductor, and the gap could be tuned by a perpendicular displacement field.

Inspired by the experimental feasibility of tuning the band structures of TLG, researchers started to investigate the relationship between the external displacement field and electronic properties. M. F. Craciun \textit{et al.} found that the field-induced interlayer asymmetry led to hybridization of the linear and parabolic parts of the band structure in ABA-TLG\cite{Craciun2009}. This led to increments of the DOS, Fermi velocity and minimal conductivity. Compared with that in TBG, the response to an external field in ABA-TLG is significant.

ABC-TLG has more practical properties compared with ABA-TLG. It was predicted by many groups that an external field induces opening of a band gap\cite{PhysRevB.79.035421,PhysRevB.80.195401,PhysRevB.81.125304,doi:10.1063/1.3595335,doi:10.1063/1.3604019,doi:10.1021/jp201761p}. Approximately two years after the first prediction, experimental research proved this. C. H. Lui \textit{et al.} monitored the band gap opening of ABC-TLG through infrared conductivity measurements, and they found that a large band gap of $\sim$120 meV was induced by a gate voltage of 1.2 V\cite{Lui2011}. W. Bao \textit{et al.} measured the relationship between the conductance and back gate voltage via standard lock-in techniques, and they found that an insulating state emerges in ABC-TLG near the charge neutrality point at low temperatures $\sim$1.5K\cite{Bao2011}. They attributed the appearance of the insulating phase to the strong electronic interactions that give rise to spontaneous symmetry breaking. This is similar to the phenomenon reported in BLG\cite{PhysRevB.82.115124,PhysRevB.81.041401,PhysRevB.81.041402,Weitz812}. The realization of a continuously tunable band gap indicates the potential application of ABC-TLG in graphene electronics and photonics, such as tunable terahertz light sources and detectors.

The substrates of multilayer graphene also influence the intrinsic properties of graphene heterostructures. The most common substrate for graphene is SiO$_2$, on which the device quality is improved\cite{PhysRevLett.101.096802,Du2008}. However, there are several limitations, such as the limited carrier mobility caused by scattering\cite{doi:10.1143/JPSJ.75.074716,106,107,108,109,110,111,112}, and carrier inhomogeneity caused by substrate-induced disorder. These limitations cause the characteristics of the samples to fall below expectations\cite{RevModPhys.81.109,106,107,108,109,110,111,112,113,114}. Therefore, researchers sought alternatives to $SiO_2$ for a long time, and in 2010, hBN became the most qualified candidate. hBN has a smooth surface without dangling bonds and charge traps, as well as a large electrical band gap of $\sim$5.97 eV and a lattice constant similar to that of graphene\cite{115}. In addition, the boron atoms and nitrogen atoms occupy A and B sublattices, respectively, just as the carbon atoms do in graphene. However, the lattice mismatch leads to moir\'{e} patterns, which were observed in 2011 by using scanning tunnelling microscopy\cite{116}. Only with a mechanical transfer process can the atoms in graphene align with the atoms in hBN. However, owing to the different chemical potentials on the different sublattices in hBN, the symmetry of the sublattices in graphene could be partially broken. This leads to emergence of a band gap\cite{117,118,119}.

\begin{figure}[tbp]
	\centering
	\includegraphics[width=0.36\textwidth]{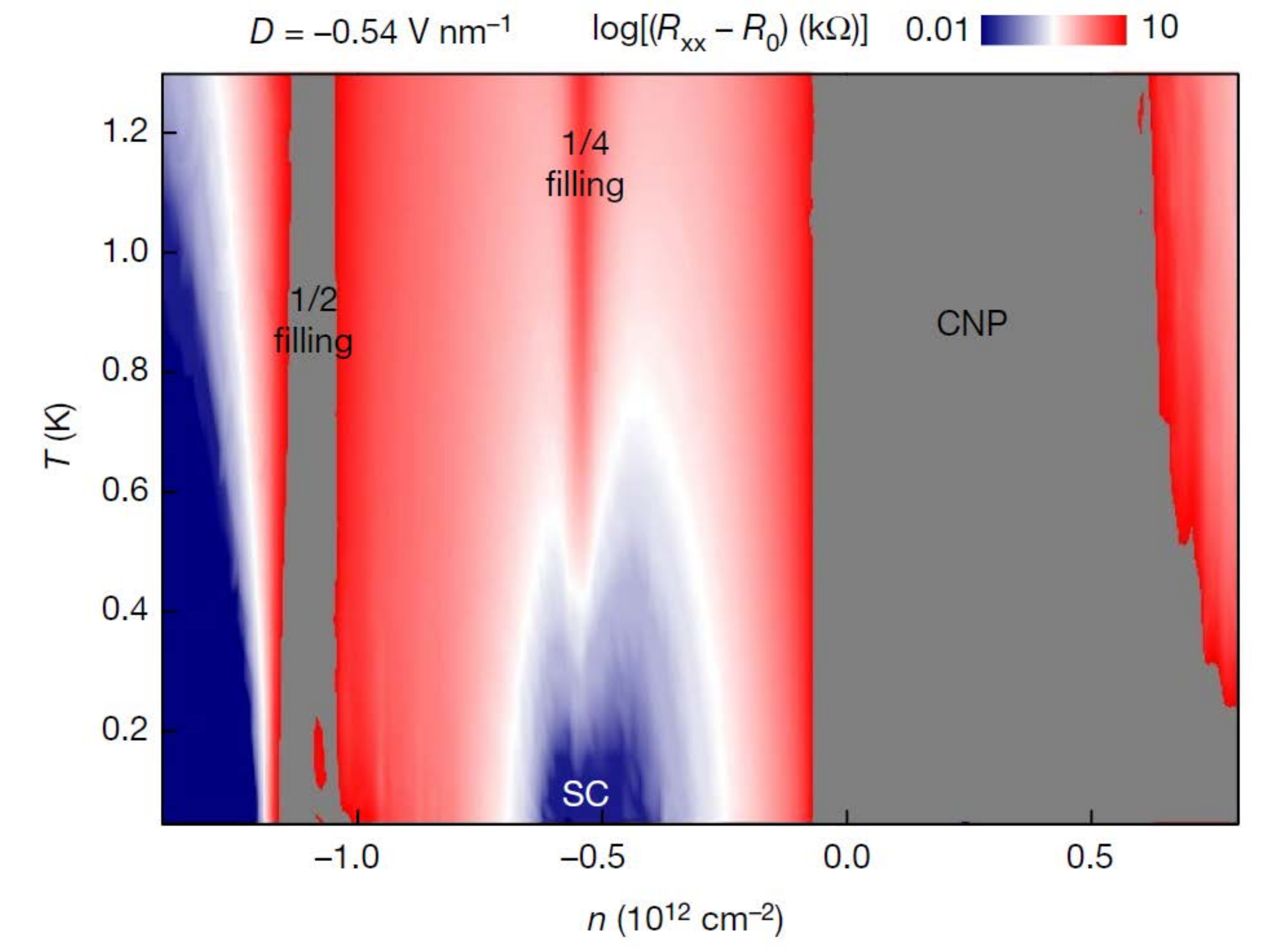}
	\caption{Carrier-density-dependent phase diagram. R$_{xx}$ represents the gate-dependent four-probe resistance, a function of the carrier density and temperature at $D=-0.54V/nm$. $R_0=380\Omega$. Reproduced with permission from Ref.\cite{124}.}
	\label{phase}
\end{figure}

Inspired by the discovery that the tBLG/hBN moir\'{e} superlattice gives rise to a superconducting-insulating transition at the magic angle, researchers started to explore similar phenomena in the TLG/hBN structure. High temperature superconductivity usually appears in a doped Mott insulator\cite{120,121,122,123}. In 2019, G. Chen \textit{et al.} fabricated a gate-tunable Mott insulator\cite{10.1038/s41567-018-0387-2} and observed signatures of superconductivity\cite{124} in ABC-TLG/hBN heterostructures. They designed a dual-gate device based on the ABC-TLG/hBN moir\'{e} superlattice, and the superlattice constant of the moir\'{e} pattern was 15 nm. They calculated the band structure and predicted the appearance of metallic, Mott-insulating, and superconducting states. Such a tunable quantum system has an isolated, nearly flat fourfold-degenerate valence miniband, and the electronic correlation could be controlled by an external displacement field. At $1/4$ or $1/2$ filling of the first hole-type miniband, Mott insulating states emerged, corresponding to one hole or two holes per superlattice unit cell, respectively. They qualitatively explained the appearance of the Mott state: the Coulomb repulsion energy between the conductive electrons is greater than the kinetic energy, which features a large electron mass and thus leads to a Mott insulator. In addition, by applying a proper displacement field of -0.54 nV/nm to tune the bandwidth and chilling the system to below 1 K, they found that superconductivity occurs in the $1/4$ filling state with a carrier density n of $-5.4\times 10^{11}cm^{-2}$.
As illustrated in Fig.\ref{phase}, two superconducting domes emerged on both the electron-doped and hole-doped sides of the ABC-TLG/hBN heterostructure. This behaviour is rather similar to that in high $T_c$ doped cuprates. The superconducting signatures are attributed to the charge inhomogeneity in ABC-TLG.

\section{Conclusions and Perspectives}
Graphene, which is a single-layer monoatomic material, displays rich and exotic physics. The simple structure of graphene makes it a pure and ideal system to study the physics of massless chiral Dirac fermions. It was theoretically predicted many years ago that superconductivity occurs in single-layer graphene, but it has not been observed in experiments. Nature never ceases to surprise: the fabulous experiments by Y. Cao et al. suggest that unconventional superconductivity occurs in TBG. In this review, we strive to summarize all of the important works in this field. However, the study of multilayer twisted graphene is a quite active field, and there are more works published every day. In the end, we are only able to mention a small fraction of them.

Several theoretical works indicate that the $d_{x^2-y^2}+id_{xy}$ ($d+id$) pairing symmetry, or the chiral $d$-wave, superconducting state can be realized in TBG. For trilayer graphene, although the $T_c$ of this ABC-TLG/hBN system is relatively low, the Mott insulator and doping-induced superconductivity are similar to those of doped cuprates. This may provide a platform to study strongly correlated electron-electron interactions. Many other novel electronic phases may also appear in the ABC-TLG/hBN heterostructure, such as the Chern insulator\cite{125}, quantum anomalous Hall effect related to electrically tunable Chern flat bands\cite{126,127,128}, topological superconductivity\cite{PhysRevLett.121.087001}, and Mott transition between a metallic Fermi liquid phase and a spin liquid insulating phase\cite{129}. It is worth mentioning about the recent proposal for emulating the physics of twisted bilayer materials beyond materials research \cite{PRL125030504,arXiv200802854}, where two synthetic layers are produced by exploiting coherently coupled internal atomic states of ultracold atoms trapped in an optical lattice and moir{\'e}-like patterns are directly imprinted on the lattice by spatially modulating the interlayer Raman coupling, without the need of a physical twist of the layers. Although many works have been performed, the pairing symmetry of the superconductivity and the mechanism of the unconventional superconductivity remain unclear.

\noindent
\underline{\it Acknowledgements} ---
This work was supported by NSFC (Nos. 11774033 and 11974049) and Beijing Natural Science Foundation (No. 1192011).
The numerical simulations in this work were performed at HSCC of
Beijing Normal University and Tianhe in the Beijing Computational Science Research Center.

\end{document}